\documentclass{WileyASNA-v1}

\usepackage{graphicx}
\usepackage{amsmath}
\usepackage{url}
\usepackage[utf8]{inputenc}
\usepackage[T1]{fontenc}

\articletype{Research Article} % article type as defined by class

\received{xx xxx 2025}
\revised{xx xxx 2025}
\accepted{xx xxx 2025}

\title{Obscured Star Formation in the Dwarf Galaxy DDO~43? A Comparative UV--IR Analysis}

\author[1]{Áron Juhász}
\author[2]{Enikő Pichler}

\authormark{Juhász et al.}

\address[1]{\orgdiv{University of Debrecen, Institute of Physics}}
\address[2]{\orgdiv{Institute of Physics and Astronomy, Eötvös Loránd University, H-1117 Budapest, Pázmány P. s. 1/A, Hungary}}

\corres{Á. Juhász, \email{juhasza92@mailbox.unideb.hu}} % <- adjust later if needed

\abstract{
We present a study of recent star formation in the dwarf irregular galaxy DDO~43 using GALEX FUV and WISE NIR imaging. We identify regions of elevated FUV flux, indicating unobscured star-forming activity across much of the galaxy. To further characterize the stellar content, we compare the FUV fluxes to archival WISE W1 and W2 infrared data across 56 regions of interest. A general correlation is found between the FUV and infrared fluxes, suggesting spatially coherent star formation throughout the galaxy. A few regions, however, show elevated infrared fluxes but little or no UV emission, potentially indicating localized, dust-obscured star formation.
}


\begin{document}

\maketitle

\section{Introduction}

DDO~43 is a nearby, gas-rich dwarf irregular galaxy located in relative isolation, with no known companion galaxies in its immediate vicinity. Its low-mass nature, irregular morphology, and lack of recent interactions make it an ideal testbed for investigating the internal processes that govern star formation. 

The star formation rate (SFR) in galaxies may be estimated in various ways \citep[see reviews by][]{Kennicutt1998, KennicuttEvans2012}: based on the UV surface brightness (e.g. \citealt{salim2007}); on the far-infrared brightness (e.g. \citealt{Heraudeau2004}); on the total infrared luminosity (e.g. \citealt{liu2020}); from the brightness of PAH lines (e.g. \citealt{kovacs2019}); or using SED modelling (e.g. \citealt{suleiman2022}).

Recent analysis by \citet{Subramanian2024} reports a stellar mass of $2.1 \times 10^7$~M$_\odot$ and an H\,\textsc{i} mass of $1.7 \times 10^8$~M$_\odot$ for DDO~43, implying a gas fraction of 0.91. The far-ultraviolet luminosity corresponds to a star formation rate of $1.8 \times 10^{-3}$~M$_\odot$\,yr$^{-1}$, and the authors identify 18 star-forming clumps with sizes ranging from 100 to 250~pc. These values are consistent with other gas-rich dwarf irregular galaxies showing distributed and clumpy star formation.

DDO~43 exhibits ongoing, spatially extended star formation activity \citep{Simpson2005}. This makes it particularly valuable for studying how local physical conditions can sustain or regulate star formation. 

Previous multiwavelength surveys, such as LITTLE~THINGS \citep{Hunter2012} and LVL \citep{Dale2009}, have shown that combining UV and IR observations provides a more complete picture of star formation activity, revealing both unobscured young stellar populations and embedded regions obscured by dust. These data are critical for identifying both the structural morphology and possible dust-obscured regions of star formation.

\begin{table*}[b]
  \centering
  \caption{Magnitude comparison for selected foreground stars. Negative differences indicate fainter IRAF values, implying overestimated background.}
  \begin{tabular}{clcccccc}
    \hline
    ID & Source ID & W1 Catalog & W2 Catalog & W1 IRAF & W2 IRAF & W1 Diff. & W2 Diff. \\
    \hline
    A & 1127p408\_b0-000918 & 12.619 & 12.713 & 12.937 & 12.940 & --0.318 & --0.227 \\
    B & 1127p408\_b0-002631 & 14.004 & 14.156 & 14.280 & 14.661 & --0.276 & --0.505 \\
    C & 1127p408\_b0-004038 & 14.593 & 14.690 & 14.790 & 15.066 & --0.197 & --0.376 \\
    D & 1127p408\_b0-005810 & 15.102 & 14.949 & 15.386 & 16.217 & --0.284 & --1.268 \\
    \hline
  \end{tabular}
  \label{tab:mag_comparison}
\end{table*}

\begin{table*}[t]
  \centering
  \caption{Total fluxes measured with IRAF (W1, W2) and fluxes implied by catalog magnitudes (W1 (cat), W2 (cat)), given in digital numbers (DN). The last four columns list the background levels (DN) and aperture areas (pix) used to estimate how much background would need to be subtracted (uniformly) to match the catalog fluxes.}
  \begin{tabular}{lcccccccc}
    \hline
    ID & W1 (DN) & W2 (DN) & W1 (cat, DN) & W2 (cat, DN) & BG$_\mathrm{W1}$ (DN) & BG$_\mathrm{W2}$ (DN) & Ap$_\mathrm{W1}$ (pix) & Ap$_\mathrm{W2}$ (pix) \\
    \hline
    A & 1524.409 & 1812.791 & 1420.365 & 518.561 & 0.867 & 8.492 & 119.944 & 152.407 \\
    B & 745.497  & 1493.289 & 396.643  & 137.278 & 2.915 & 8.906 & 119.682 & 152.266 \\
    C & 634.626  & 1490.135 & 230.569  & 83.946  & 3.375 & 9.207 & 119.720 & 152.726 \\
    D & 549.390  & 1456.439 & 144.278  & 66.130  & 3.375 & 9.103 & 120.027 & 152.735 \\
    \hline
  \end{tabular}
  \label{tab:flux_comparison}
\end{table*}

Motivated by these findings, we apply this combined approach to DDO~43 using archival far-ultraviolet (FUV) data from the Galaxy Evolution Explorer (GALEX; \citep{Martin2005}) and infrared imaging from the Wide-field Infrared Survey Explorer (WISE; \citep{Wright2010}). Specifically, we analyze the W1 (3.4~$\mu$m) and W2 (4.6~$\mu$m) bands alongside FUV data to assess the galaxy’s star-forming morphology and the presence of potentially obscured regions. Our goal is to evaluate whether the UV and IR tracers yield a consistent picture of star formation in this system.

\section{Data and Methods}

\subsection{Infrared Imaging and Background Treatment}
\label{sec:infrared}

The infrared data were obtained from the WISE W1 (3.4~$\mu$m) and W2 (4.6~$\mu$m) bands \citep{Cutri2013}, which primarily trace the Rayleigh–Jeans (RJ) tail of the stellar continuum emitted by evolved, low-mass stars \citep{Jarrett2023}. As such, both W1 and W2 are predominantly sensitive to the older stellar population, and their luminosities serve as proxies for stellar mass. The images have angular resolutions of 6.1" in W1 and 6.4" in W2 \citep{Wright2010}, enabling moderately resolved surface brightness measurements in nearby dwarf galaxies.

W2 is also sensitive to additional emission components such as warm dust, and can therefore exhibit subtle excesses in dusty or star-forming environments \citep{simonian2017}.

While direct tracers of thermal dust emission are typically found in longer-wavelength bands such as W3 (12~$\mu$m) and W4 (22~$\mu$m) — which include prominent PAH features \citep{Nikutta2014} — DDO~43 remains undetected in these bands due to its low infrared surface brightness and strong background fluctuations. We therefore rely on W1 and W2 imaging, using the W1--W2 color as a proxy for identifying potential warm dust emission \citep{Stern2012}.

\begin{figure}[t]
  \centering
  \includegraphics[width=\columnwidth]{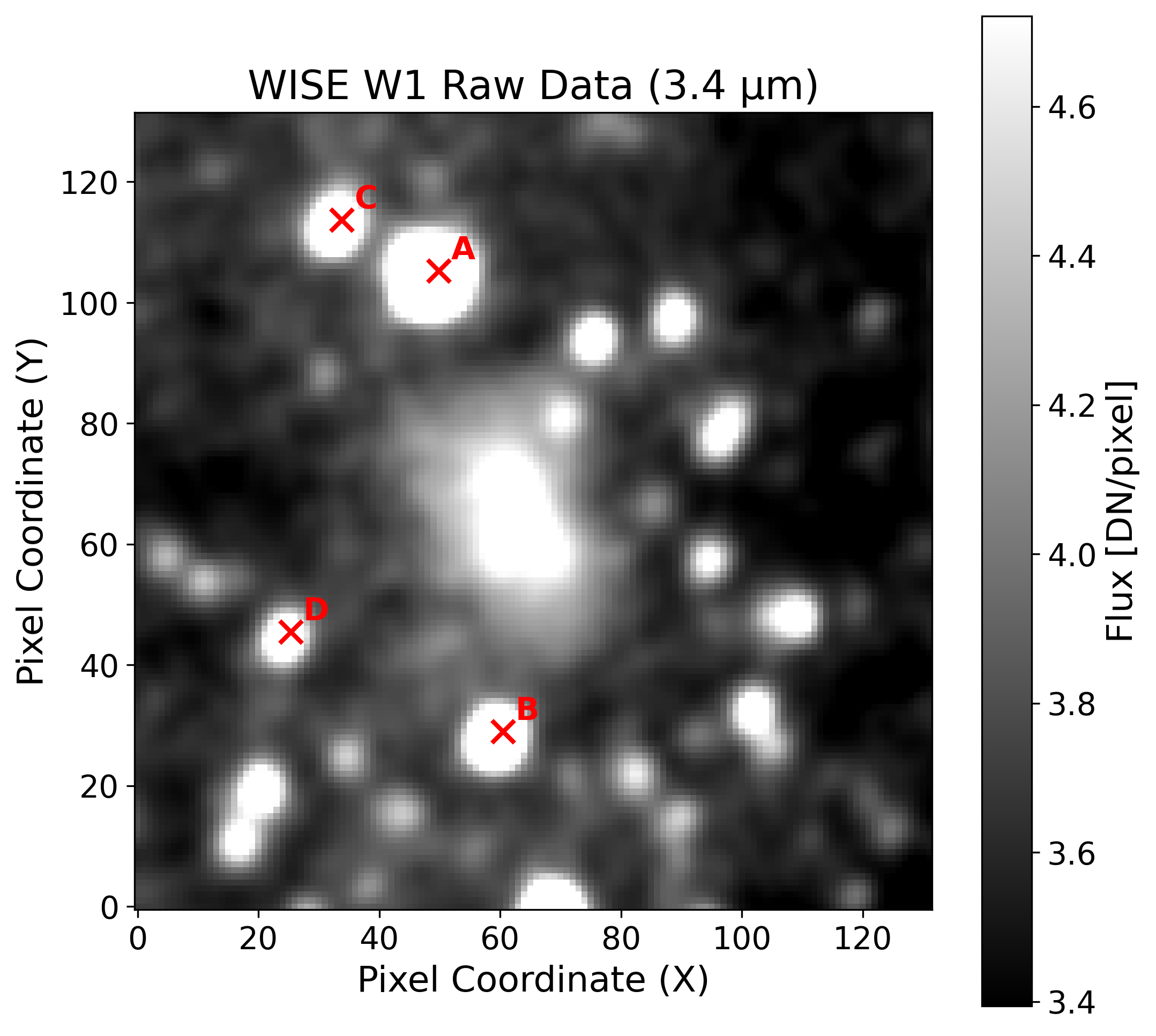}
  \caption{WISE W1 image (3.4 $\mu$m) showing the positions of the four foreground stars (A–D) used for photometric comparison. The galaxy center lies slightly southeast of star A. The selected stars broadly sample the surroundings of the galaxy, allowing for background estimation across different local conditions.}
  \label{w1_stars}
\end{figure}

\begin{figure*}[t]
  \centering
  \includegraphics[width=\textwidth]{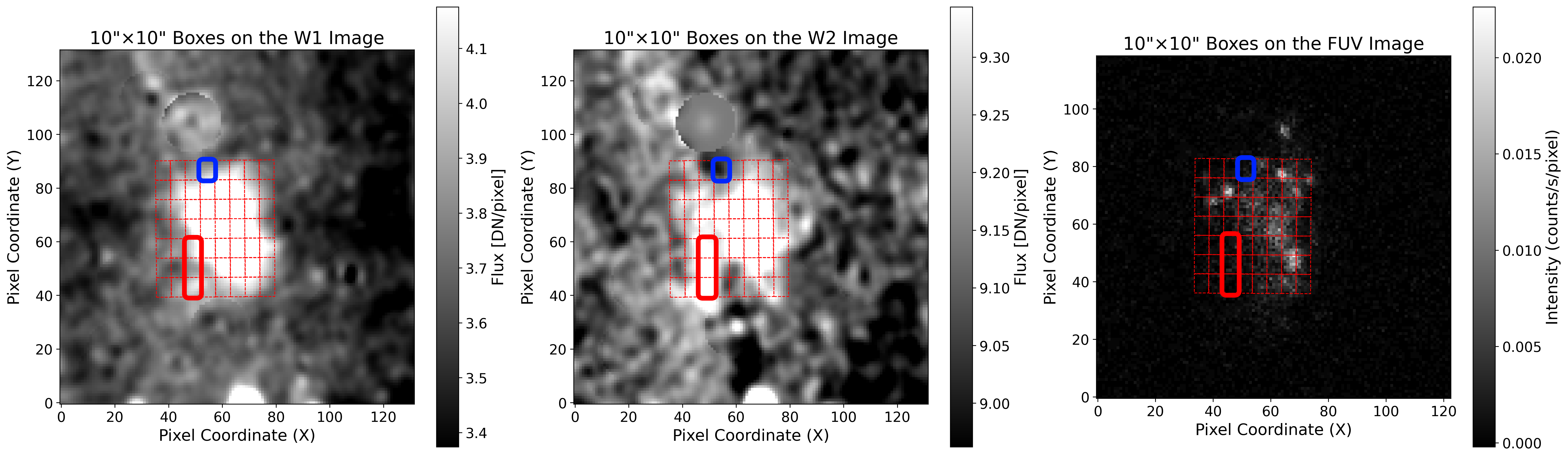}
  \caption{W1 (left), W2 (middle), and FUV (right) images with overlaid 10$\times$10 arcsec$^2$ sampling boxes. Red frames indicate W2-bright, UV-faint regions (SE1-3), corresponding to dust-enshrouded star-forming clumps in the southeastern quadrant. The blue frame marks a W1-bright, W2- and UV-faint region in the north (N1), likely dominated by an evolved stellar population.}
  \label{w1_w2_uv}
\end{figure*}

The images were retrieved from the official WISE data release, where the pipeline-processed frames are already calibrated and corrected for instrumental artifacts. Pixel values are given in digital numbers (DN) per pixel, which can be converted into flux densities and subsequently Vega-system magnitudes using the standard conversion formulae provided in the WISE documentation \citep{Cutri2013}.

The WISE images of DDO~43 are dominated by foreground stars \citep{Lang2014} from the Milky Way and are known to suffer from background fluctuations. In the W2 band, after foreground star subtraction, the brightest pixel within the galaxy reaches a digital number (DN) value of $\sim$9.8, while the local background is $\sim$9.0. In W1, the central peak reaches $\sim$5.24 DN against a background level of $\sim$3.4. Expressed as simple contrasts, these correspond to $\sim$12\% in W2 and $\sim$60\% in W1 relative to the background level. However, the relevant metric for detection significance is the signal-to-noise ratio. 

To estimate the background noise amplitude, we first removed contaminating foreground stars through PSF photometry, and then measured statistics in the cleaned edge strips of the images. This procedure yields $\sigma \approx 0.07$\,DN\,pix$^{-1}$ in both W1 and W2. In W1, the net signal above the background (1.83 DN) corresponds to $\sim$26$\sigma$, while in W2 the net signal (0.8 DN) corresponds to $\sim$11$\sigma$. These values confirm that the galaxy is robustly detected in both bands, even though the overall contrast relative to the background level is modest. Accurate background subtraction and noise characterization remain critical for any meaningful photometric analysis.

In addition, the W2 mosaic exhibits a weak, nearly vertical dark stripe running in the northeast--southwest direction, effectively slicing through the image and also crossing the galaxy. While the stripe is not strong enough to dominate the photometry, it represents an instrumental artifact that must be kept in mind in later sections.

Our goal was to measure the surface brightness in selected regions of the galaxy. The background is shaped by a combination of diffuse Galactic emission \citep{Schlegel1998}, interplanetary dust scattering \citep{Kelsall1998}, and instrumental effects. Unlike in the GALEX data, where the sky level could be directly adopted, here the presence of structured fluctuations and instrumental artifacts meant that a simple global background estimate would not have been reliable. Accurate modeling of all these components would be necessary to derive a scientifically defensible background level, but this exceeds the scope and capabilities of our dataset. Therefore, instead of attempting a detailed physical decomposition, we adopted a more practical solution.

To mitigate these challenges, we employed a targeted background correction strategy based on IRAF aperture photometry. Several foreground stars were selected and measured in both W1 and W2 bands (see Figure~\ref{w1_stars}). By comparing their instrumental magnitudes to catalog values from the CatWISE2020 catalog \citep{Marocco2021}, we estimated the effective background contribution within typical photometric apertures and applied a uniform correction accordingly (see Table~\ref{tab:mag_comparison} and Table~\ref{tab:flux_comparison}, and Figure~\ref{w1_stars}). This method allows us to recover approximate flux measurements under suboptimal conditions.

\subsection{Ultraviolet Imaging}

We analyzed archival FUV imaging of DDO~43 from the GALEX space telescope. The GALEX FUV channel provides a spatial resolution of 4.5" \citep{Martin2005}, which is sharper than that of the WISE W1 and W2 bands. In our data, the sky background—measured via sigma-clipped statistics across the full frame—has a mean level of approximately $2.2\times10^{-4}$ counts s$^{-1}$ pix$^{-1}$, with a pixel-to-pixel rms noise of $3.1\times10^{-4}$ counts s$^{-1}$ pix$^{-1}$. The brightest clumps reach peak brightness levels of $\sim$0.02 counts s$^{-1}$ pix$^{-1}$, corresponding to a signal-to-noise ratio of $\sim$60--70. Even outside these brightest areas, the galaxy’s diffuse FUV emission typically exceeds the background by a factor of 10 to 50, allowing for robust photometric analysis throughout most of the disk. 

The FUV background is exceptionally low and spatially uniform \citep{Morrissey2007}, across the field, largely free from the structured contamination. In this wavelength regime, there are no significant contributions from interplanetary dust \citep{Bianchi2011}, diffuse Galactic emission, or instrumental artifacts. 

The GALEX data are pipeline-processed and calibrated, with pixel values given in counts per second per pixel \citep{Morrissey2007}. These values can be converted into flux densities and AB magnitudes using the provided calibration constants. While we did not carry out a full photometric conversion for this study, we relied on the native count-rate values for relative surface brightness comparisons.

To accurately determine the sky background level, we applied a sigma-clipping technique to the entire image. Specifically, we performed a $3\sigma$ iterative clipping with a maximum of five iterations, producing robust estimates of mean, median, and standard deviation. This allowed us to statistically characterize the background level while minimizing the influence of outliers. Since the GALEX data are largely free from the structured artifacts and large-scale fluctuations that affect the WISE images, the resulting background value could be directly adopted for the surface brightness measurements of the galaxy.

\begin{figure*}[t]
  \centering
  \includegraphics[width=\textwidth]{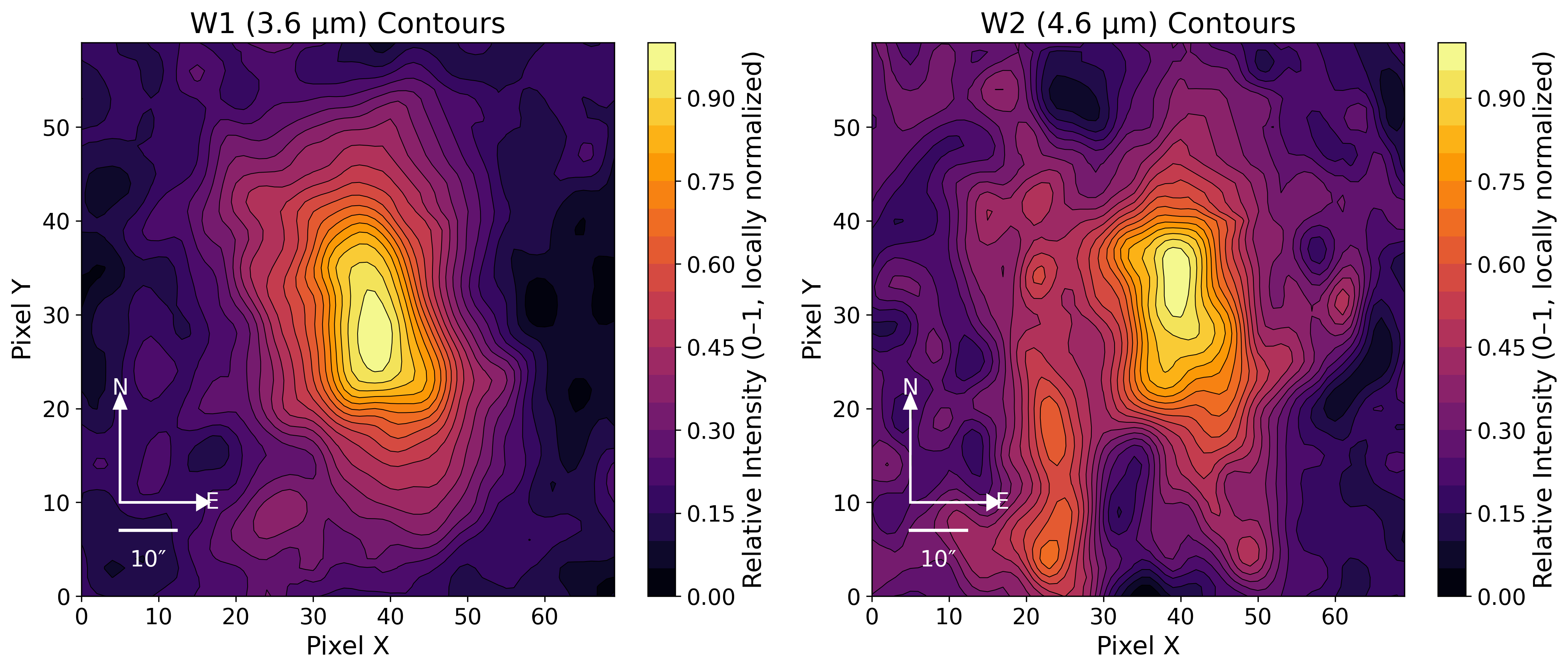}
  \caption{Locally normalized intensity contour maps for W1 (left) and W2 (right). The W1 image shows a regular, centrally concentrated profile, while the W2 emission is more asymmetric, with an extended structure toward the southeast. A faint vertical artifact in W2 (running northwest--southeast) passes through the galaxy and may visually enhance the apparent separation of the southeastern clumps. Taking this into account, the overall morphology still supports the interpretation of localized dust-obscured star formation.}
  \label{contours}
\end{figure*}

In addition, isolated zero-valued pixels were present within the galaxy, likely due to detector artifacts or local sensitivity issues. To avoid artificially underestimating the surface brightness in these regions, we replaced the affected pixels with the mean of their immediate, non-zero neighbors. This correction preserved the overall morphology while allowing for a more reliable statistical analysis of the galaxy's emission.

\subsection{UV--IR Sampling Strategy}

To compare ultraviolet and infrared emission on a consistent spatial basis, we defined 56 identical sampling regions across the galaxy. The sampling layout was guided by a H\,\textsc{i} analysis of DDO~43 \citep{Pichler2025b}. Each region is a $10\times10$~arcsec box, corresponding to approximately 380~pc at the assumed distance of DDO~43. The boxes were arranged to cover the central part of the galaxy. The angular resolution of the data varies between bands: GALEX FUV provides $\sim$4.5" full width at half maximum (FWHM; \citep{Martin2005}), while WISE W1 and W2 have resolutions of 6.1" and 6.4", respectively \citep{Wright2010}. All of these values are significantly smaller than the $10\times10$~arcsec sampling boxes, ensuring that the resolution differences do not critically affect the interpretation of the flux measurements within each region.

Care was taken to exclude areas strongly contaminated by residual foreground stars or background artifacts, as identified in earlier processing steps. In particular, regions surrounding the bright subtracted stars used for background calibration (sources A and B; see Figure~\ref{w1_stars} and Table~\ref{tab:mag_comparison}) often display PSF-related residual structures that could distort local flux measurements. The final grid was therefore restricted to the main body of the galaxy, deliberately avoiding these visually compromised peripheral zones.

For each valid region, we measured the background-corrected FUV, W1, and W2 fluxes, and computed W1--W2 color indices in the Vega magnitude system. This spatially matched sampling provides a foundation for assessing the degree of correlation and disparity between ultraviolet and infrared emission, and for identifying regions of potentially obscured star formation. While some peripheral zones appear redder, even these regions remain under the AGN threshold (W1--W2 $<$ 0.8; \citep{Stern2012}), and none of the redder zones are centrally located.

\section{Results}

\subsection{Morphology}

 The FUV image reveals several bright clumps, particularly concentrated in the northern and eastern portions of the galaxy (see third panel of Figure~\ref{w1_w2_uv}). These regions are likely associated with recent, unobscured star formation, possibly representing OB associations or fragmented stellar complexes.

In comparison, the WISE W1 image displays a relatively regular, mildly elliptical shape with an orientation from the northwest to the southeast. This suggests a symmetric near-infrared stellar light distribution. The W2 image, by contrast, appears more irregular and less elliptical. It features several localized protrusions, and in the southeastern quadrant, a spur-like extension appears partially separated from the main body of the galaxy. These morphological differences between W1 and W2 (see Figure~\ref{contours}) highlight structural asymmetries that are not visible in the UV or W1 images.

\subsection{General UV--IR Correlation}

\begin{figure*}[b]
  \centering
  \includegraphics[width=\textwidth]{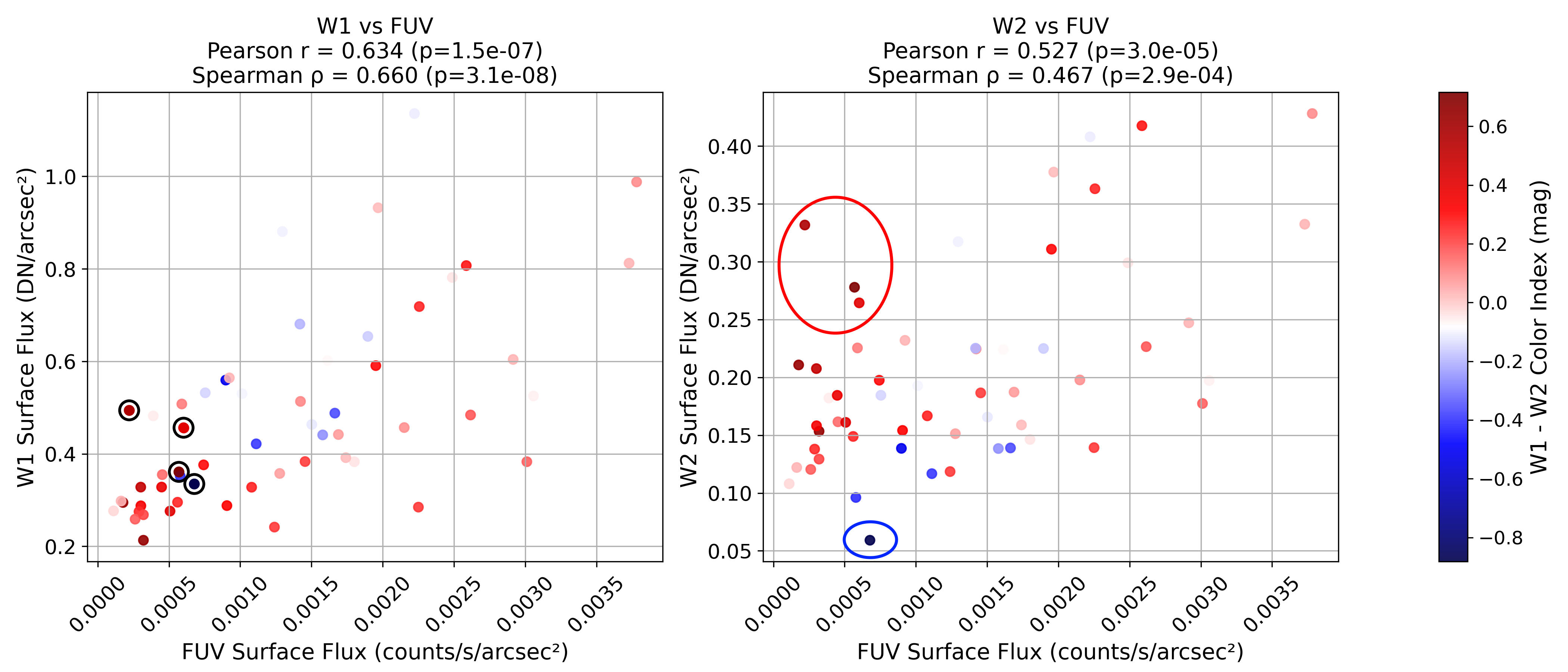}
  \caption{Correlation between GALEX FUV surface brightness and WISE W1 (left) and W2 (right). Points are color-coded by the W1$-$W2 color index (redder values indicate warmer dust emission). Pearson and Spearman correlation coefficients are shown above each panel. While the correlation is stronger in W1, consistent with relatively unobscured star formation, the W2--FUV plot reveals several outliers: three W2-bright, UV-dark red points (circled in red) and one blue point with low W2 and no UV emission (circled in blue). Excluding these, the W2--FUV correlation would likely be comparable to that of W1. For reference, unreddened main-sequence stars typically have W1$-$W2 $\approx$ 0\,mag, since their spectra follow the Rayleigh--Jeans tail in this wavelength regime. Normal star-forming galaxies, on the other hand, generally fall in the range W1$-$W2 $\approx$ 0--0.5\,mag \citep{Stern2012}, which provides context for assessing the unusually red colors of the W2-bright regions.}
  \label{uv_vs_ir}
\end{figure*}

To explore the relationship between unobscured and dust-embedded star formation, we compared the measured fluxes in the GALEX FUV, WISE W1, and W2 bands across the 56 sampled regions (see Figure~\ref{uv_vs_ir}). As expected, a positive correlation is observed between the FUV and infrared fluxes in most areas of the galaxy. Regions that are bright in FUV tend to show elevated W1 and W2 emission as well, consistent with spatially coincident young stellar populations and their associated stellar mass and dust content.
\begin{table*}[t]
\centering
\caption{Surface brightness and color index for selected regions.}
\label{tab:SE_outliers}
\begin{tabular}{lcccccc}
\hline
Region & W1 Flux & W1 (mag) & W2 Flux & W2 (mag) & W1--W2 & FUV Flux \\
& (DN / arcsec$^2$) & (Vega) & (DN / arcsec$^2$) & (Vega) & (mag) & (counts s$^{-1}$ arcsec$^{-2}$) \\
\hline
SE1 & 0.456 & 21.35 & 0.264 & 20.94 & 0.41 & 0.00060 \\
SE2 & 0.361 & 21.61 & 0.278 & 20.89 & 0.72 & 0.00057 \\
SE3 & 0.494 & 21.27 & 0.332 & 20.70 & 0.57 & 0.00022 \\
N1  & 0.335 & 21.69 & 0.059 & 22.57 & $-0.88$ & 0.00068 \\
\hline
\end{tabular}
\end{table*}
Quantitatively, the FUV–W1 correlation shows high significance, with Pearson’s $r = 0.634$ and Spearman’s $\rho = 0.660$. The FUV–W2 correlation is somewhat weaker ($r = 0.527$, $\rho = 0.467$), primarily due to a few outlier regions, discussed below.

\subsection{Outlier Regions and Dust Obscuration}

A subset of regions in the southeastern quadrant deviates markedly from the overall trend. These outliers exhibit strong W2 emission but minimal or no FUV signal. In these cases, the W1 flux is also relatively weak, suggesting that the emission is not simply due to old stellar populations. Rather, the elevated W2 values combined with suppressed ultraviolet flux point to localized star-forming regions that are heavily obscured by dust, rendering them invisible in the UV (see Figure~\ref{w1_w2_uv} and Table~\ref{tab:SE_outliers}).

Importantly, these W2-bright clumps are located just to the west of the weak vertical artifact noted in Sect.~\ref{sec:infrared}. Their fluxes are therefore not directly contaminated by the stripe itself. Nevertheless, the presence of the artifact means that the separation of these clumps from the central body of the galaxy (as shown in Fig.~\ref{contours}) might appear more pronounced than it really is. We therefore treat their interpretation with caution, while still regarding them as plausible candidates for dust-enshrouded star formation.

In contrast, we identified one additional outlier in the northern region with a notably blue W1--W2 color index, showing significant W1 flux but negligible W2 and FUV emission. This region likely reflects a passively evolving stellar population without active star formation or warm dust — effectively the opposite of the southeastern dusty clumps (see Figure~\ref{w1_w2_uv} and Table~\ref{tab:SE_outliers}).

Excluding these four outliers — three W2-bright and one W2-faint — the remaining 52 sampled regions exhibit broadly consistent behavior. FUV, W1, and W2 fluxes are moderately correlated throughout, reinforcing the overall picture of a galaxy dominated by spatially coherent star formation, with only localized departures due to dust obscuration or population gradients.

\section{Discussion and Conclusions}

Our analysis of ultraviolet and infrared imaging of DDO~43 reveals a predominantly consistent picture of spatially coherent star formation throughout the galaxy, punctuated by a small number of anomalous regions. The general correlation between FUV, W1, and W2 fluxes supports the view that most star formation in this system is unobscured and moderately active.

However, the presence of several W2-bright, UV-dark regions — particularly in the southeastern quadrant — points to a localized component of dust-enshrouded star formation. These areas display elevated W2 flux without a corresponding increase in UV emission, and their red W1--W2 color indices are inconsistent with passively evolving stellar populations. In contrast, one region with a blue W1--W2 index and no FUV emission likely reflects a passively evolving stellar population, devoid of both young stars and warm dust \citep[e.g.,][]{Jarrett2023}. The coexistence of such varied environments within a single, relatively small dwarf galaxy underscores the complex internal structure of DDO~43.

In general, W1 traces the older stellar population, while W2 includes contributions from both evolved stars and warm dust emission, including very small grains \citep{simonian2017}. Typical W1--W2 colors for normal star-forming galaxies range between 0 and 0.5mag, with values exceeding 0.8mag generally associated with AGN activity \citep{Stern2012}. In our sample, the reddest regions (e.g., SE2, SE3) exhibit W1--W2 values of 0.57--0.72mag (see Table\ref{tab:SE_outliers}), placing them near the upper range expected for dust-rich star-forming clumps but still below the AGN threshold. This further supports their interpretation as localized regions with enhanced warm dust emission rather than non-stellar contamination.

Minor contributions from stellar population effects such as metallicity-dependent CO absorption may also affect the W1--W2 color \citep{Norris2014}, although we do not model this explicitly here. Red W1--W2 colors may also arise from AGN activity in dwarf galaxies \citep{Aravindan2024}, but this is unlikely in our case given that all redder regions remain below the AGN threshold and are located away from the galactic center.

The spatial offset between the infrared and ultraviolet emission — particularly the asymmetry observed in W2 — raises the possibility of non-uniform star formation triggers. While DDO~43 is classified as an isolated system, the presence of a morphologically distinct, dust-rich zone may point to past internal instabilities or subtle external influences, such as minor tidal interactions \citep{Subramanian2024} or gas accretion events. These scenarios remain speculative in the absence of kinematic data, but highlight the value of multiwavelength analysis in revealing structural and evolutionary features that may otherwise go unnoticed.

Similar combinations of obscured and unobscured regions have been reported in other low-mass dwarf irregular galaxies. \citet{Zhang2012} found that a significant fraction of such systems exhibit “outside-in” star formation histories, where recent activity becomes increasingly confined to the central regions, while the outer disk becomes quiescent. The presence of both UV-bright, W2-faint zones and W2-bright, UV-dark clumps within DDO~43 fits well into this broader picture, suggesting that this galaxy may represent a typical example of the heterogeneous internal structure and evolutionary path seen among low-mass dwarfs.

In summary, our results suggest that DDO~43 hosts a primarily unobscured, moderately active star-forming population, with localized regions of potential embedded star formation and internal asymmetry. Narrow-band H$\alpha$ imaging from the LITTLE~THINGS survey shows no pronounced emission at the location of the W2-bright region; this neither proves nor rules out embedded star formation, since very dusty environments could remain invisible even in H$\alpha$. Future studies incorporating spatially resolved H\,\textsc{i} data and mid-infrared spectroscopy may help further clarify the origin and nature of these features. Optical and infrared spectroscopy and imaging with high spatial resolution — for example, using SUBARU/FOCAS \citep{Kashikawa2002} or SUBARU/IRCS — could both reveal kinematic signatures of past perturbations and directly test whether the W2-bright, UV-dark region indeed hosts heavily obscured star formation. We are preparing a proposal for such SUBARU/IRCS observations, since the $L$- and $M$-bands are closely matched to the W1 and W2 wavelength regime. Likewise, deeper and higher-resolution ultraviolet imaging, such as with \textit{UVIT} onboard the \textit{AstroSat} mission \citep{Tandon2017}, could confirm whether the W2-bright, UV-dark regions are truly lacking in UV emission or merely unresolved in the current GALEX data.

\section*{Acknowledgements}

Supported by the University of Debrecen Program for Scientific Publication.

Based on observations made with the NASA Galaxy Evolution Explorer. GALEX is operated for NASA by the California Institute of Technology under NASA contract NAS5-98034.

This publication makes use of data products from the Wide-field Infrared Survey Explorer, which is a joint project of the University of California, Los Angeles, and the Jet Propulsion Laboratory/California Institute of Technology, funded by the National Aeronautics and Space Administration.

We are grateful to Deidre Hunter and Bruce Elmegreen for their helpful comments on the manuscript and for their leadership in the LITTLE THINGS project, which included DDO~43. 

We also thank the anonymous referee for the many constructive comments and suggestions that greatly improved the paper.

\bibliography{Wiley-ASNA}

\begin{thebibliography}{}

\bibitem [\protect \citeauthoryear {%
Bettonvil%
, Sutterlin%
, Hammerschlag%
, Rutten%
\BCBL {}\ \BBA {} Stix%
}{%
Bettonvil%
\ \protect \BOthers {.}}{%
{\protect \APACyear {2003}}%
}]{%
Bettonvil2003}
\APACinsertmetastar {%
Bettonvil2003}%
\begin{APACrefauthors}%
Bettonvil, F\BPBI C.%
, Sutterlin, P.%
, Hammerschlag, R\BPBI H.%
, Rutten, R\BPBI J.%
\BCBL {}\ \BBA {} Stix, M.%
\end{APACrefauthors}%
\unskip\
\newblock
\APACrefYearMonthDay{2003}{}{},
\newblock
{\BBOQ}\APACrefatitle {Proc. SPIE Conf. Ser.} {Proc. SPIE Conf. Ser.}{\BBCQ}
\newblock
\BIn{} \BVOL\ 4853, \BPG~306.
\PrintBackRefs{\CurrentBib}

\bibitem [\protect \citeauthoryear {%
Bland-Hawthorn%
, van Breugel%
\BCBL {}\ \BBA {} Gillingham%
}{%
Bland-Hawthorn%
\ \protect \BOthers {.}}{%
{\protect \APACyear {2001}}%
}]{%
Bland2001}
\APACinsertmetastar {%
Bland2001}%
\begin{APACrefauthors}%
Bland-Hawthorn, J.%
, van Breugel, W.%
\BCBL {}\ \BBA {} Gillingham, P\BPBI R.%
\end{APACrefauthors}%
\unskip\
\newblock
\APACrefYearMonthDay{2001}{}{},
\newblock
\unskip
\newblock
\APACjournalVolNumPages{ApJ}{563}{}{611}.
\PrintBackRefs{\CurrentBib}

\bibitem [\protect \citeauthoryear {%
Kosugi%
\ \BBA {} Gillingham%
}{%
Kosugi%
\ \BBA {} Gillingham%
}{%
{\protect \APACyear {2007}}%
}]{%
Kosugi2007}
\APACinsertmetastar {%
Kosugi2007}%
\begin{APACrefauthors}%
Kosugi, T.%
\BCBT {}\ \BBA {} Gillingham, R\BPBI H.%
\end{APACrefauthors}%
\unskip\
\newblock
\APACrefYearMonthDay{2007}{}{},
\newblock
\unskip
\newblock
\APACjournalVolNumPages{Sol. Phys.}{243}{}{3}.
\PrintBackRefs{\CurrentBib}

\bibitem [\protect \citeauthoryear {%
Kosugi%
\ \protect \BOthers {.}}{%
Kosugi%
\ \protect \BOthers {.}}{%
{\protect \APACyear {2009}}%
}]{%
Kosugi2009}
\APACinsertmetastar {%
Kosugi2009}%
\begin{APACrefauthors}%
Kosugi, T.%
, Matsuzaki, K.%
, Sakao, R.%
, Bettonvil, F\BPBI C.%
, Sutterlin, P.%
\BCBL {}\ \BBA {} Hammerschlag, R\BPBI H.%
\end{APACrefauthors}%
\unskip\
\newblock
\APACrefYearMonthDay{2009}{}{},
\newblock
\unskip
\newblock
\APACjournalVolNumPages{Sol. Phys.}{243}{}{3}.
\PrintBackRefs{\CurrentBib}

\bibitem [\protect \citeauthoryear {%
Power%
\ \protect \BOthers {.}}{%
Power%
\ \protect \BOthers {.}}{%
{\protect \APACyear {1975}}%
}]{%
Paivio1975}
\APACinsertmetastar {%
Paivio1975}%
\begin{APACrefauthors}%
Power, J\BPBI D.%
, Cohen, A\BPBI L.%
, Nelson, S\BPBI M.%
\ et al.\end{APACrefauthors}%
\unskip\
\newblock
\APACrefYearMonthDay{1975}{}{},
\newblock
\unskip
\newblock
\APACjournalVolNumPages{Cognition}{37}{2}{635}.
\PrintBackRefs{\CurrentBib}

\bibitem [\protect \citeauthoryear {%
Rutten%
}{%
Rutten%
}{%
{\protect \APACyear {2007}}%
}]{%
Rutten2007}
\APACinsertmetastar {%
Rutten2007}%
\begin{APACrefauthors}%
Rutten, R\BPBI J.%
\end{APACrefauthors}%
\unskip\
\newblock
\APACrefYearMonthDay{2007}{}{},
\newblock
{\BBOQ}\APACrefatitle {The Physics of Chromospheric Plasmas} {The Physics of
  Chromospheric Plasmas}.{\BBCQ}
\newblock
\BIn{} P.~Heinzel, I.~Dorotovic\BCBL {}\ \BBA {} R\BPBI J.~Rutten\ (\BEDS),
  \APACrefbtitle {ASP Conf. Ser.} {ASP Conf. Ser.}\ \BVOL~368, \BPG~27.
\PrintBackRefs{\CurrentBib}

\bibitem [\protect \citeauthoryear {%
Stix%
}{%
Stix%
}{%
{\protect \APACyear {2004}}%
}]{%
Stix2004}
\APACinsertmetastar {%
Stix2004}%
\begin{APACrefauthors}%
Stix, M.%
\end{APACrefauthors}%
\unskip\
\newblock
\APACrefYear{2004},
\newblock
\APACrefbtitle {Astronomy and Astrophysics Library} {Astronomy and Astrophysics
  Library}\ (\PrintOrdinal{2}\ \BEd).
\newblock
\APACaddressPublisher{Berlin}{Springer}.
\PrintBackRefs{\CurrentBib}

\bibitem [\protect \citeauthoryear {%
{Strunk Jr.}%
\ \BBA {} White%
}{%
{Strunk Jr.}%
\ \BBA {} White%
}{%
{\protect \APACyear {1979}}%
}]{%
Strunk1979}
\APACinsertmetastar {%
Strunk1979}%
\begin{APACrefauthors}%
{Strunk Jr.}, W.%
\BCBT {}\ \BBA {} White, E\BPBI B.%
\end{APACrefauthors}%
\unskip\
\newblock
\APACrefYear{1979},
\newblock
\APACrefbtitle {The Elements of Style} {The Elements of Style}\
  (\PrintOrdinal{3}\ \BEd).
\newblock
\APACaddressPublisher{New York}{MacMillan}.
\PrintBackRefs{\CurrentBib}

\end{thebibliography}


\begin{thebibliography}{}

\bibitem [\protect \citeauthoryear {%
{Aravindan}%
, {Canalizo}%
, {Secrest}%
, {Satyapal}%
\BCBL {}\ \BBA {} {Bohn}%
}{%
{Aravindan}%
\ \protect \BOthers {.}}{%
{\protect \APACyear {2024}}%
}]{%
Aravindan2024}
\APACinsertmetastar {%
Aravindan2024}%
\begin{APACrefauthors}%
{Aravindan}, A.%
, {Canalizo}, G.%
, {Secrest}, N.%
, {Satyapal}, S.%
\BCBL {}\ \BBA {} {Bohn}, T.%
\end{APACrefauthors}%
\unskip\
\newblock
\APACrefYearMonthDay{2024}{{\APACmonth{11}}}{},
\newblock
\unskip
\newblock
\APACjournalVolNumPages{\apj}{975}{1}{60}.
\newblock
\begin{APACrefDOI} \doi{10.3847/1538-4357/ad702b} \end{APACrefDOI}
\PrintBackRefs{\CurrentBib}

\bibitem [\protect \citeauthoryear {%
{Bianchi}%
\ \protect \BOthers {.}}{%
{Bianchi}%
\ \protect \BOthers {.}}{%
{\protect \APACyear {2011}}%
}]{%
Bianchi2011}
\APACinsertmetastar {%
Bianchi2011}%
\begin{APACrefauthors}%
{Bianchi}, L.%
, {Herald}, J.%
, {Efremova}, B.%
\ et al.\end{APACrefauthors}%
\unskip\
\newblock
\APACrefYearMonthDay{2011}{{\APACmonth{09}}}{},
\newblock
\unskip
\newblock
\APACjournalVolNumPages{\apss}{335}{1}{161-169}.
\newblock
\begin{APACrefDOI} \doi{10.1007/s10509-010-0581-x} \end{APACrefDOI}
\PrintBackRefs{\CurrentBib}

\bibitem [\protect \citeauthoryear {%
{Cutri}%
\ \protect \BOthers {.}}{%
{Cutri}%
\ \protect \BOthers {.}}{%
{\protect \APACyear {2021}}%
}]{%
Cutri2013}
\APACinsertmetastar {%
Cutri2013}%
\begin{APACrefauthors}%
{Cutri}, R\BPBI M.%
, {Wright}, E\BPBI L.%
, {Conrow}, T.%
\ et al.\end{APACrefauthors}%
\unskip\
\newblock
\APACrefYearMonthDay{2021}{{\APACmonth{02}}}{},
\newblock
\APACrefbtitle {{VizieR Online Data Catalog: AllWISE Data Release (Cutri+
  2013)}.} {{VizieR Online Data Catalog: AllWISE Data Release (Cutri+ 2013)}.},
\newblock
\APAChowpublished {VizieR On-line Data Catalog: II/328. Originally published
  in: IPAC/Caltech (2013)}.
\PrintBackRefs{\CurrentBib}

\bibitem [\protect \citeauthoryear {%
{Dale}%
\ \protect \BOthers {.}}{%
{Dale}%
\ \protect \BOthers {.}}{%
{\protect \APACyear {2009}}%
}]{%
Dale2009}
\APACinsertmetastar {%
Dale2009}%
\begin{APACrefauthors}%
{Dale}, D\BPBI A.%
, {Cohen}, S\BPBI A.%
, {Johnson}, L\BPBI C.%
\ et al.\end{APACrefauthors}%
\unskip\
\newblock
\APACrefYearMonthDay{2009}{{\APACmonth{09}}}{},
\newblock
\unskip
\newblock
\APACjournalVolNumPages{\apj}{703}{1}{517-556}.
\newblock
\begin{APACrefDOI} \doi{10.1088/0004-637X/703/1/517} \end{APACrefDOI}
\PrintBackRefs{\CurrentBib}

\bibitem [\protect \citeauthoryear {%
{H{\'e}raudeau}%
\ \protect \BOthers {.}}{%
{H{\'e}raudeau}%
\ \protect \BOthers {.}}{%
{\protect \APACyear {2004}}%
}]{%
Heraudeau2004}
\APACinsertmetastar {%
Heraudeau2004}%
\begin{APACrefauthors}%
{H{\'e}raudeau}, P.%
, {Oliver}, S.%
, {del Burgo}, C.%
\ et al.\end{APACrefauthors}%
\unskip\
\newblock
\APACrefYearMonthDay{2004}{{\APACmonth{11}}}{},
\newblock
\unskip
\newblock
\APACjournalVolNumPages{\mnras}{354}{3}{924-934}.
\newblock
\begin{APACrefDOI} \doi{10.1111/j.1365-2966.2004.08259.x} \end{APACrefDOI}
\PrintBackRefs{\CurrentBib}

\bibitem [\protect \citeauthoryear {%
{Hunter}%
\ \protect \BOthers {.}}{%
{Hunter}%
\ \protect \BOthers {.}}{%
{\protect \APACyear {2012}}%
}]{%
Hunter2012}
\APACinsertmetastar {%
Hunter2012}%
\begin{APACrefauthors}%
{Hunter}, D\BPBI A.%
, {Ficut-Vicas}, D.%
, {Ashley}, T.%
\ et al.\end{APACrefauthors}%
\unskip\
\newblock
\APACrefYearMonthDay{2012}{{\APACmonth{11}}}{},
\newblock
\unskip
\newblock
\APACjournalVolNumPages{\aj}{144}{5}{134}.
\newblock
\begin{APACrefDOI} \doi{10.1088/0004-6256/144/5/134} \end{APACrefDOI}
\PrintBackRefs{\CurrentBib}

\bibitem [\protect \citeauthoryear {%
{Jarrett}%
\ \protect \BOthers {.}}{%
{Jarrett}%
\ \protect \BOthers {.}}{%
{\protect \APACyear {2023}}%
}]{%
Jarrett2023}
\APACinsertmetastar {%
Jarrett2023}%
\begin{APACrefauthors}%
{Jarrett}, T\BPBI H.%
, {Cluver}, M\BPBI E.%
, {Taylor}, E\BPBI N.%
, {Bellstedt}, S.%
, {Robotham}, A\BPBI S\BPBI G.%
\BCBL {}\ \BBA {} {Yao}, H\BPBI F\BPBI M.%
\end{APACrefauthors}%
\unskip\
\newblock
\APACrefYearMonthDay{2023}{{\APACmonth{04}}}{},
\newblock
\unskip
\newblock
\APACjournalVolNumPages{\apj}{946}{2}{95}.
\newblock
\begin{APACrefDOI} \doi{10.3847/1538-4357/acb68f} \end{APACrefDOI}
\PrintBackRefs{\CurrentBib}

\bibitem [\protect \citeauthoryear {%
{Kashikawa}%
\ \protect \BOthers {.}}{%
{Kashikawa}%
\ \protect \BOthers {.}}{%
{\protect \APACyear {2002}}%
}]{%
Kashikawa2002}
\APACinsertmetastar {%
Kashikawa2002}%
\begin{APACrefauthors}%
{Kashikawa}, N.%
, {Aoki}, K.%
, {Asai}, R.%
\ et al.\end{APACrefauthors}%
\unskip\
\newblock
\APACrefYearMonthDay{2002}{{\APACmonth{12}}}{},
\newblock
\unskip
\newblock
\APACjournalVolNumPages{\pasj}{54}{6}{819-832}.
\newblock
\begin{APACrefDOI} \doi{10.1093/pasj/54.6.819} \end{APACrefDOI}
\PrintBackRefs{\CurrentBib}

\bibitem [\protect \citeauthoryear {%
{Kelsall}%
\ \protect \BOthers {.}}{%
{Kelsall}%
\ \protect \BOthers {.}}{%
{\protect \APACyear {1998}}%
}]{%
Kelsall1998}
\APACinsertmetastar {%
Kelsall1998}%
\begin{APACrefauthors}%
{Kelsall}, T.%
, {Weiland}, J\BPBI L.%
, {Franz}, B\BPBI A.%
\ et al.\end{APACrefauthors}%
\unskip\
\newblock
\APACrefYearMonthDay{1998}{{\APACmonth{11}}}{},
\newblock
\unskip
\newblock
\APACjournalVolNumPages{\apj}{508}{1}{44-73}.
\newblock
\begin{APACrefDOI} \doi{10.1086/306380} \end{APACrefDOI}
\PrintBackRefs{\CurrentBib}

\bibitem [\protect \citeauthoryear {%
R\BPBI C.~{Kennicutt}%
\ \BBA {} {Evans}%
}{%
R\BPBI C.~{Kennicutt}%
\ \BBA {} {Evans}%
}{%
{\protect \APACyear {2012}}%
}]{%
KennicuttEvans2012}
\APACinsertmetastar {%
KennicuttEvans2012}%
\begin{APACrefauthors}%
{Kennicutt}, R\BPBI C.%
\BCBT {}\ \BBA {} {Evans}, N\BPBI J.%
\end{APACrefauthors}%
\unskip\
\newblock
\APACrefYearMonthDay{2012}{{\APACmonth{09}}}{},
\newblock
\unskip
\newblock
\APACjournalVolNumPages{\araa}{50}{}{531-608}.
\newblock
\begin{APACrefDOI} \doi{10.1146/annurev-astro-081811-125610} \end{APACrefDOI}
\PrintBackRefs{\CurrentBib}

\bibitem [\protect \citeauthoryear {%
R\BPBI C.~{Kennicutt} Jr.%
}{%
R\BPBI C.~{Kennicutt} Jr.%
}{%
{\protect \APACyear {1998}}%
}]{%
Kennicutt1998}
\APACinsertmetastar {%
Kennicutt1998}%
\begin{APACrefauthors}%
{Kennicutt}, R\BPBI C., Jr.%
\end{APACrefauthors}%
\unskip\
\newblock
\APACrefYearMonthDay{1998}{{\APACmonth{01}}}{},
\newblock
\unskip
\newblock
\APACjournalVolNumPages{\araa}{36}{}{189-232}.
\newblock
\begin{APACrefDOI} \doi{10.1146/annurev.astro.36.1.189} \end{APACrefDOI}
\PrintBackRefs{\CurrentBib}

\bibitem [\protect \citeauthoryear {%
{Kov{\'a}cs}%
\ \protect \BOthers {.}}{%
{Kov{\'a}cs}%
\ \protect \BOthers {.}}{%
{\protect \APACyear {2019}}%
}]{%
kovacs2019}
\APACinsertmetastar {%
kovacs2019}%
\begin{APACrefauthors}%
{Kov{\'a}cs}, T\BPBI O.%
, {Burgarella}, D.%
, {Kaneda}, H.%
, {Moln{\'a}r}, D\BPBI C.%
, {Oyabu}, S.%
, {Pinter}, S.%
\BCBL {}\ \BBA {} {Toth}, L\BPBI V.%
\end{APACrefauthors}%
\unskip\
\newblock
\APACrefYearMonthDay{2019}{{\APACmonth{04}}}{},
\newblock
\unskip
\newblock
\APACjournalVolNumPages{\pasj}{71}{2}{27}.
\newblock
\begin{APACrefDOI} \doi{10.1093/pasj/psy145} \end{APACrefDOI}
\PrintBackRefs{\CurrentBib}

\bibitem [\protect \citeauthoryear {%
{Lang}%
}{%
{Lang}%
}{%
{\protect \APACyear {2014}}%
}]{%
Lang2014}
\APACinsertmetastar {%
Lang2014}%
\begin{APACrefauthors}%
{Lang}, D.%
\end{APACrefauthors}%
\unskip\
\newblock
\APACrefYearMonthDay{2014}{{\APACmonth{05}}}{},
\newblock
\unskip
\newblock
\APACjournalVolNumPages{\aj}{147}{5}{108}.
\newblock
\begin{APACrefDOI} \doi{10.1088/0004-6256/147/5/108} \end{APACrefDOI}
\PrintBackRefs{\CurrentBib}

\bibitem [\protect \citeauthoryear {%
{Liu}%
\ \protect \BOthers {.}}{%
{Liu}%
\ \protect \BOthers {.}}{%
{\protect \APACyear {2020}}%
}]{%
liu2020}
\APACinsertmetastar {%
liu2020}%
\begin{APACrefauthors}%
{Liu}, T.%
, {Evans}, N\BPBI J.%
, {Kim}, K\BHBI T.%
\ et al.\end{APACrefauthors}%
\unskip\
\newblock
\APACrefYearMonthDay{2020}{{\APACmonth{08}}}{},
\newblock
\unskip
\newblock
\APACjournalVolNumPages{\mnras}{496}{3}{2821-2835}.
\newblock
\begin{APACrefDOI} \doi{10.1093/mnras/staa1501} \end{APACrefDOI}
\PrintBackRefs{\CurrentBib}

\bibitem [\protect \citeauthoryear {%
{Marocco}%
\ \protect \BOthers {.}}{%
{Marocco}%
\ \protect \BOthers {.}}{%
{\protect \APACyear {2021}}%
}]{%
Marocco2021}
\APACinsertmetastar {%
Marocco2021}%
\begin{APACrefauthors}%
{Marocco}, F.%
, {Eisenhardt}, P\BPBI R\BPBI M.%
, {Fowler}, J\BPBI W.%
\ et al.\end{APACrefauthors}%
\unskip\
\newblock
\APACrefYearMonthDay{2021}{{\APACmonth{03}}}{},
\newblock
\unskip
\newblock
\APACjournalVolNumPages{\apjs}{253}{1}{8}.
\newblock
\begin{APACrefDOI} \doi{10.3847/1538-4365/abd805} \end{APACrefDOI}
\PrintBackRefs{\CurrentBib}

\bibitem [\protect \citeauthoryear {%
{Martin}%
\ \protect \BOthers {.}}{%
{Martin}%
\ \protect \BOthers {.}}{%
{\protect \APACyear {2005}}%
}]{%
Martin2005}
\APACinsertmetastar {%
Martin2005}%
\begin{APACrefauthors}%
{Martin}, D\BPBI C.%
, {Fanson}, J.%
, {Schiminovich}, D.%
\ et al.\end{APACrefauthors}%
\unskip\
\newblock
\APACrefYearMonthDay{2005}{{\APACmonth{01}}}{},
\newblock
\unskip
\newblock
\APACjournalVolNumPages{\apjl}{619}{1}{L1-L6}.
\newblock
\begin{APACrefDOI} \doi{10.1086/426387} \end{APACrefDOI}
\PrintBackRefs{\CurrentBib}

\bibitem [\protect \citeauthoryear {%
{Morrissey}%
\ \protect \BOthers {.}}{%
{Morrissey}%
\ \protect \BOthers {.}}{%
{\protect \APACyear {2007}}%
}]{%
Morrissey2007}
\APACinsertmetastar {%
Morrissey2007}%
\begin{APACrefauthors}%
{Morrissey}, P.%
, {Conrow}, T.%
, {Barlow}, T\BPBI A.%
\ et al.\end{APACrefauthors}%
\unskip\
\newblock
\APACrefYearMonthDay{2007}{{\APACmonth{12}}}{},
\newblock
\unskip
\newblock
\APACjournalVolNumPages{\apjs}{173}{2}{682-697}.
\newblock
\begin{APACrefDOI} \doi{10.1086/520512} \end{APACrefDOI}
\PrintBackRefs{\CurrentBib}

\bibitem [\protect \citeauthoryear {%
{Nikutta}%
, {Hunt-Walker}%
, {Nenkova}%
, {Ivezi{\'c}}%
\BCBL {}\ \BBA {} {Elitzur}%
}{%
{Nikutta}%
\ \protect \BOthers {.}}{%
{\protect \APACyear {2014}}%
}]{%
Nikutta2014}
\APACinsertmetastar {%
Nikutta2014}%
\begin{APACrefauthors}%
{Nikutta}, R.%
, {Hunt-Walker}, N.%
, {Nenkova}, M.%
, {Ivezi{\'c}}, {\v{Z}}.%
\BCBL {}\ \BBA {} {Elitzur}, M.%
\end{APACrefauthors}%
\unskip\
\newblock
\APACrefYearMonthDay{2014}{{\APACmonth{08}}}{},
\newblock
\unskip
\newblock
\APACjournalVolNumPages{\mnras}{442}{4}{3361-3379}.
\newblock
\begin{APACrefDOI} \doi{10.1093/mnras/stu1087} \end{APACrefDOI}
\PrintBackRefs{\CurrentBib}

\bibitem [\protect \citeauthoryear {%
{Norris}%
\ \protect \BOthers {.}}{%
{Norris}%
\ \protect \BOthers {.}}{%
{\protect \APACyear {2014}}%
}]{%
Norris2014}
\APACinsertmetastar {%
Norris2014}%
\begin{APACrefauthors}%
{Norris}, M\BPBI A.%
, {Meidt}, S.%
, {Van de Ven}, G.%
, {Schinnerer}, E.%
, {Groves}, B.%
\BCBL {}\ \BBA {} {Querejeta}, M.%
\end{APACrefauthors}%
\unskip\
\newblock
\APACrefYearMonthDay{2014}{{\APACmonth{12}}}{},
\newblock
\unskip
\newblock
\APACjournalVolNumPages{\apj}{797}{1}{55}.
\newblock
\begin{APACrefDOI} \doi{10.1088/0004-637X/797/1/55} \end{APACrefDOI}
\PrintBackRefs{\CurrentBib}

\bibitem [\protect \citeauthoryear {%
Pichler%
, Koncz%
, Gabanyi%
, Joó%
\BCBL {}\ \BBA {} Tóth%
}{%
Pichler%
\ \protect \BOthers {.}}{%
{\protect \APACyear {2025}}%
}]{%
Pichler2025b}
\APACinsertmetastar {%
Pichler2025b}%
\begin{APACrefauthors}%
Pichler, E.%
, Koncz, B.%
, Gabanyi, K\BPBI E.%
, Joó, A\BPBI P.%
\BCBL {}\ \BBA {} Tóth, L\BPBI V.%
\end{APACrefauthors}%
\unskip\
\newblock
\APACrefYearMonthDay{2025}{Mar.}{},
\newblock
\unskip
\newblock
\APACjournalVolNumPages{Acta Polytechnica}{65}{1}{73–78}.
\newblock
\begin{APACrefURL} \url{https://ojs.cvut.cz/ojs/index.php/ap/article/view/9973}
  \end{APACrefURL}
\newblock
\begin{APACrefDOI} \doi{10.14311/AP.2025.65.0073} \end{APACrefDOI}
\PrintBackRefs{\CurrentBib}

\bibitem [\protect \citeauthoryear {%
{Salim}%
\ \protect \BOthers {.}}{%
{Salim}%
\ \protect \BOthers {.}}{%
{\protect \APACyear {2007}}%
}]{%
salim2007}
\APACinsertmetastar {%
salim2007}%
\begin{APACrefauthors}%
{Salim}, S.%
, {Rich}, R\BPBI M.%
, {Charlot}, S.%
\ et al.\end{APACrefauthors}%
\unskip\
\newblock
\APACrefYearMonthDay{2007}{{\APACmonth{12}}}{},
\newblock
\unskip
\newblock
\APACjournalVolNumPages{\apjs}{173}{2}{267-292}.
\newblock
\begin{APACrefDOI} \doi{10.1086/519218} \end{APACrefDOI}
\PrintBackRefs{\CurrentBib}

\bibitem [\protect \citeauthoryear {%
{Schlegel}%
, {Finkbeiner}%
\BCBL {}\ \BBA {} {Davis}%
}{%
{Schlegel}%
\ \protect \BOthers {.}}{%
{\protect \APACyear {1998}}%
}]{%
Schlegel1998}
\APACinsertmetastar {%
Schlegel1998}%
\begin{APACrefauthors}%
{Schlegel}, D\BPBI J.%
, {Finkbeiner}, D\BPBI P.%
\BCBL {}\ \BBA {} {Davis}, M.%
\end{APACrefauthors}%
\unskip\
\newblock
\APACrefYearMonthDay{1998}{{\APACmonth{06}}}{},
\newblock
\unskip
\newblock
\APACjournalVolNumPages{\apj}{500}{2}{525-553}.
\newblock
\begin{APACrefDOI} \doi{10.1086/305772} \end{APACrefDOI}
\PrintBackRefs{\CurrentBib}

\bibitem [\protect \citeauthoryear {%
{Simonian}%
\ \BBA {} {Martini}%
}{%
{Simonian}%
\ \BBA {} {Martini}%
}{%
{\protect \APACyear {2017}}%
}]{%
simonian2017}
\APACinsertmetastar {%
simonian2017}%
\begin{APACrefauthors}%
{Simonian}, G\BPBI V.%
\BCBT {}\ \BBA {} {Martini}, P.%
\end{APACrefauthors}%
\unskip\
\newblock
\APACrefYearMonthDay{2017}{{\APACmonth{02}}}{},
\newblock
\unskip
\newblock
\APACjournalVolNumPages{\mnras}{464}{4}{3920-3936}.
\newblock
\begin{APACrefDOI} \doi{10.1093/mnras/stw2623} \end{APACrefDOI}
\PrintBackRefs{\CurrentBib}

\bibitem [\protect \citeauthoryear {%
{Simpson}%
, {Hunter}%
\BCBL {}\ \BBA {} {Nordgren}%
}{%
{Simpson}%
\ \protect \BOthers {.}}{%
{\protect \APACyear {2005}}%
}]{%
Simpson2005}
\APACinsertmetastar {%
Simpson2005}%
\begin{APACrefauthors}%
{Simpson}, C\BPBI E.%
, {Hunter}, D\BPBI A.%
\BCBL {}\ \BBA {} {Nordgren}, T\BPBI E.%
\end{APACrefauthors}%
\unskip\
\newblock
\APACrefYearMonthDay{2005}{{\APACmonth{09}}}{},
\newblock
\unskip
\newblock
\APACjournalVolNumPages{\aj}{130}{3}{1049-1064}.
\newblock
\begin{APACrefDOI} \doi{10.1086/432537} \end{APACrefDOI}
\PrintBackRefs{\CurrentBib}

\bibitem [\protect \citeauthoryear {%
{Stern}%
\ \protect \BOthers {.}}{%
{Stern}%
\ \protect \BOthers {.}}{%
{\protect \APACyear {2012}}%
}]{%
Stern2012}
\APACinsertmetastar {%
Stern2012}%
\begin{APACrefauthors}%
{Stern}, D.%
, {Assef}, R\BPBI J.%
, {Benford}, D\BPBI J.%
\ et al.\end{APACrefauthors}%
\unskip\
\newblock
\APACrefYearMonthDay{2012}{{\APACmonth{07}}}{},
\newblock
\unskip
\newblock
\APACjournalVolNumPages{\apj}{753}{1}{30}.
\newblock
\begin{APACrefDOI} \doi{10.1088/0004-637X/753/1/30} \end{APACrefDOI}
\PrintBackRefs{\CurrentBib}

\bibitem [\protect \citeauthoryear {%
{Subramanian}%
, {Mondal}%
\BCBL {}\ \BBA {} {Kalari}%
}{%
{Subramanian}%
\ \protect \BOthers {.}}{%
{\protect \APACyear {2024}}%
}]{%
Subramanian2024}
\APACinsertmetastar {%
Subramanian2024}%
\begin{APACrefauthors}%
{Subramanian}, S.%
, {Mondal}, C.%
\BCBL {}\ \BBA {} {Kalari}, V.%
\end{APACrefauthors}%
\unskip\
\newblock
\APACrefYearMonthDay{2024}{{\APACmonth{01}}}{},
\newblock
\unskip
\newblock
\APACjournalVolNumPages{\aap}{681}{}{A8}.
\newblock
\begin{APACrefDOI} \doi{10.1051/0004-6361/202346536} \end{APACrefDOI}
\PrintBackRefs{\CurrentBib}

\bibitem [\protect \citeauthoryear {%
{Suleiman}%
\ \protect \BOthers {.}}{%
{Suleiman}%
\ \protect \BOthers {.}}{%
{\protect \APACyear {2022}}%
}]{%
suleiman2022}
\APACinsertmetastar {%
suleiman2022}%
\begin{APACrefauthors}%
{Suleiman}, N.%
, {Noboriguchi}, A.%
, {Toba}, Y.%
\ et al.\end{APACrefauthors}%
\unskip\
\newblock
\APACrefYearMonthDay{2022}{{\APACmonth{10}}}{},
\newblock
\unskip
\newblock
\APACjournalVolNumPages{\pasj}{74}{5}{1157-1185}.
\newblock
\begin{APACrefDOI} \doi{10.1093/pasj/psac061} \end{APACrefDOI}
\PrintBackRefs{\CurrentBib}

\bibitem [\protect \citeauthoryear {%
{Tandon}%
\ \protect \BOthers {.}}{%
{Tandon}%
\ \protect \BOthers {.}}{%
{\protect \APACyear {2017}}%
}]{%
Tandon2017}
\APACinsertmetastar {%
Tandon2017}%
\begin{APACrefauthors}%
{Tandon}, S\BPBI N.%
, {Subramaniam}, A.%
, {Girish}, V.%
\ et al.\end{APACrefauthors}%
\unskip\
\newblock
\APACrefYearMonthDay{2017}{{\APACmonth{09}}}{},
\newblock
\unskip
\newblock
\APACjournalVolNumPages{\aj}{154}{3}{128}.
\newblock
\begin{APACrefDOI} \doi{10.3847/1538-3881/aa8451} \end{APACrefDOI}
\PrintBackRefs{\CurrentBib}

\bibitem [\protect \citeauthoryear {%
{Wright}%
\ \protect \BOthers {.}}{%
{Wright}%
\ \protect \BOthers {.}}{%
{\protect \APACyear {2010}}%
}]{%
Wright2010}
\APACinsertmetastar {%
Wright2010}%
\begin{APACrefauthors}%
{Wright}, E\BPBI L.%
, {Eisenhardt}, P\BPBI R\BPBI M.%
, {Mainzer}, A\BPBI K.%
\ et al.\end{APACrefauthors}%
\unskip\
\newblock
\APACrefYearMonthDay{2010}{{\APACmonth{12}}}{},
\newblock
\unskip
\newblock
\APACjournalVolNumPages{\aj}{140}{6}{1868-1881}.
\newblock
\begin{APACrefDOI} \doi{10.1088/0004-6256/140/6/1868} \end{APACrefDOI}
\PrintBackRefs{\CurrentBib}

\bibitem [\protect \citeauthoryear {%
{Zhang}%
, {Hunter}%
, {Elmegreen}%
, {Gao}%
\BCBL {}\ \BBA {} {Schruba}%
}{%
{Zhang}%
\ \protect \BOthers {.}}{%
{\protect \APACyear {2012}}%
}]{%
Zhang2012}
\APACinsertmetastar {%
Zhang2012}%
\begin{APACrefauthors}%
{Zhang}, H\BHBI X.%
, {Hunter}, D\BPBI A.%
, {Elmegreen}, B\BPBI G.%
, {Gao}, Y.%
\BCBL {}\ \BBA {} {Schruba}, A.%
\end{APACrefauthors}%
\unskip\
\newblock
\APACrefYearMonthDay{2012}{{\APACmonth{02}}}{},
\newblock
\unskip
\newblock
\APACjournalVolNumPages{\aj}{143}{2}{47}.
\newblock
\begin{APACrefDOI} \doi{10.1088/0004-6256/143/2/47} \end{APACrefDOI}
\PrintBackRefs{\CurrentBib}

\end{thebibliography}

\end{document}